\documentclass{aa}

\usepackage{txfonts,graphicx,wasysym}

\usepackage{siunitx}
\usepackage{booktabs}
\usepackage{hyperref}
\hypersetup{colorlinks=true,linkcolor=blue,filecolor=blue,urlcolor=blue,citecolor=blue}
\usepackage{xcolor}
\usepackage{comment}
\bibpunct{(}{)}{;}{a}{}{,}
\usepackage{amssymb}
\usepackage{amsmath}
\usepackage{longtable}
\usepackage{appendix}
\usepackage{lscape}
\usepackage[switch]{lineno} 

\usepackage{txfonts}
\makeatletter
\renewcommand*\aa@pageof{, page \thepage{} of \pageref*{LastPage}}
\makeatother

\usepackage{graphicx}
\begin{document}

   \title{Exploring cosmic magnetism with gamma-ray burst afterglow emission}

   \author{P. Da Vela
          \inst{1},
          D. Miceli\inst{2,3,4}, 
          L. Nava\inst{4,5},
          G. Ghirlanda\inst{4}
          %\fnmsep
          %\thanks{Just to show the usage
          %of the elements in the author field}
          }

   \institute{INAF Osservatorio di Astrofisica e Scienza dello Spazio di Bologna, Via Piero Gobetti 93/3, 40129 Bologna, Italy\\
              \email{paolo.davela@inaf.it}
         \and
             Gran Sasso Science Institute, Viale F. Crispi 7, L’Aquila (AQ), I-67100, Italy \\
             \email{davide.miceli@gssi.it}
         \and
         INFN, Sezione di Padova, Via Marzolo 8, 35131 Padova, Italy
         \and
         INAF, Osservatorio Astronomico di Brera, via Emilio Bianchi 46, 23807 Merate, Italy
         \and
         INFN, Sezione di Trieste, via Valerio 2, 34127 Trieste, Italy
             }

   \date{Received 18 December 2025; accepted 28 July 2026}
 
  \abstract
   {The nature and origin of magnetic fields on cosmological scales are still unclear. Magnetic fields detected in galaxies and galaxy clusters are typically interpreted as the result of the amplification of weak seed fields, but their nature remains largely unknown with two scenarios considered: the cosmological and the astrophysical origin. Signatures of magnetization in cosmic voids from observations of very high energy (VHE, E > 100 GeV) photons from extragalactic sources can provide crucial results. Indeed, if a non-negligible intergalactic magnetic field (IGMF) is present  in the voids, time-delayed emission known as pair-echo is expected. The timing and intensity of this signal encode information on the IGMF strength $B$ and properties. Given the recent detection of gamma-ray burst (GRB) afterglows at TeV energies, for this study we performed a detectability study of pair-echo signatures from GRBs. We simulated afterglow emission for different values of the jet kinetic energy ($E_{\rm k,iso} = 10^{49} - 10^{55}$ erg), redshift ($z = 0.03 - 1$), and light-curve break times, and estimated the expected pair-echo radiation for IGMF strengths in the range $B$ = $10^{-19} - 10^{-16}$ G. We investigated the capability of CTAO to detect the resulting emission at tens of GeV. We find that a subsample of GRBs in the $z - E_{\rm k,iso}$ parameter space can produce a detectable pair-echo component for CTAO for all the tested IGMF strengths. A steepening of the GRB afterglow light curve, caused, for example, by an early ($0.1 - 1$ days) jet break, is a key factor to increase the chances of detection. CTAO observations starting from $10-12$ h after the GRB trigger and extending up to a few days can provide valuable information on the IGMF.}

   \keywords{astroparticle physics - magnetic fields -gamma-ray burst: general - gamma rays: general
               }

   \maketitle
%
%-------------------------------------------------------------------

\section{Introduction}
Magnetic fields play a crucial role in the Universe on all scales, from stars to galaxies and galaxy clusters. However, the origin of magnetic fields in the large-scale Universe remains one of the long-standing open questions in cosmology. Although magnetic field structures have been mapped in several galaxies, the processes that generate such fields  are still largely unknown. Magnetic fields observed in galaxies are thought to have been amplified from weak seed fields through processes such as the $\alpha-\omega$ dynamo (see e.g. \citealt{Kulsrud08}) and/or compression and turbulent plasma motions during galaxy formation (\citealt{Grasso01}).

The nature of these initial seed fields is poorly understood. Two main hypotheses have been proposed for their origin --- cosmological and astrophysical \citep{Durrer13, Grasso01} --- with a mixed scenario also being considered \citep{Vazza25, Garaldi21}. The main difference between the two scenarios lies in the expected magnetization of cosmic voids. In the cosmological hypothesis, a non-negligible magnetic field is expected to pervade the intergalactic medium, including voids between galaxies. In the astrophysical hypothesis, although the ejection of magnetized material through outflows and winds could in principle seed magnetic fields beyond galaxies \citep{Furlanetto01}, significant magnetic fields are not expected to persist far from large-scale structures \citep{Ghosh25}. Therefore, searching for signatures of magnetization in regions devoid of matter (cosmic voids) is crucial to distinguish between the cosmological or astrophysical scenario. Since the intergalactic magnetic field (IGMF) is generally modelled as a stochastic, turbulent field, its macroscopic properties are defined by two key parameters: the field strength (often the root-mean-square B) and the magnetic correlation length $\lambda_B$. It has never been measured and Faraday rotation measurements limit its strength below $10^{-9}$\,G \citep{Pshirkov16}.

A particularly sensitive technique to constrain the IGMF relies on $\gamma$-ray observations of extragalactic sources. Very high energy (VHE, $E > 100$ GeV) photons interact with the extragalactic background light (EBL), producing electron–positron pairs and initiating electromagnetic cascades over cosmological distances. For instance, a 20 TeV photon has a mean-free path of order of $\sim 40$ Mpc (e.g. \citealt{Neronov09}). The resulting e$^{+}$-e$^{-}$ pairs can then upscatter cosmic microwave background (CMB) photons via inverse-Compton (IC) scattering. In this case, the mean-free path of the pairs produced by the absorption of a 20 TeV photon is of the order of 10 kpc. The IC process generates secondary $\gamma$-rays with characteristic energies $E \simeq 320 (E_0^\prime / 20,\mathrm{TeV})^2$ GeV, where $E_0^\prime$ is the energy of the primary VHE photon at the emission redshift z. These photons can further interact with the EBL, initiating further steps of the electromagnetic cascade.

If the pairs experience a non-negligible IGMF during the propagation, they are deflected, and the cascade emission becomes both morphologically extended (pair-halo emission) and temporally delayed (pair-echo emission) with respect to the original VHE source position and emission time, respectively. {In the absence of an IGMF, the cascade emission produced by the pairs occurs approximately along the line of sight. In this case, the cascade contribution is maximized; for some sources, the spectral luminosity of the cascade can exceed the intrinsic spectral luminosity of the source within the secondary $\gamma$-ray energy range, even though the total integrated luminosity of the cascade remains bounded by the primary VHE output.}

Searches for cascade signatures in blazars have yielded lower limits on the IGMF strength at the level of $\sim10^{-17}$-$10^{-15}$ G, based on the non-detection of spectral, spatial, and temporal cascade features (see e.g. \citealt{Neronov10}, \citealt{Taylor11}, \citealt{Tavecchio11}, \citealt{Dermer11}, \citealt{Finke15}, \citealt{Ackermann18}, \citealp{Podlesnyi22}, \citealt{Meyer23} and \citealt{Acciari23}).{The differences among these results stem from assumptions on the intrinsic VHE spectra and source duty cycles. Shorter duty cycles yield more conservative IGMF lower limits, as they exclude photons with time delays longer than the assumed activity periods. The most conservative approach equates the duty cycle to the VHE monitoring period: 10-15 years in early studies (\citealt{Dermer11, Taylor11, Ackermann18} and \citealt{Meyer23}), based on facility lifetimes, extending to 20 or more years today. Alternatively, some models adopt 10$^7$–10$^8$ year timescales \citep{Parma02}, producing more stringent IGMF constraints.}

Gamma-ray bursts (GRBs), being VHE emitters, can also be used to probe the IGMF (\citealt{Plaga95}, \citealt{Murase07}, \citealt{Murase08}, \citealt{Ichiki08} and \citealt{Takahashi11}). In the case of GRBs, the search for pair-echo emission follows a transient VHE event rather than a persistent blazar. The first GRB detected in the VHE band was GRB 190114C \citep{Acciari19} at redshift
$z=0.42$ \citep{Selsing19}.  The MAGIC (Major Atmospheric Gamma-ray Imaging Cherenkov Telescopes) telescopes detected VHE emission in the afterglow phase in the $0.2 - 1$ TeV range over about 40 minutes \citep{190114C_MAGIC_2019b}. While in \cite{Wang20} magnetic fields weaker than $10^{-19.5}$ G considering a correlation length less than 1 Mpc are excluded, \cite{Dzhatdoev20} and \cite{DaVela23} argue that no limits on IGMF can be placed by GRB190114C.
{Similar studies have been carried out on the most powerful GRB ever detected, GRB 221009A (z=0.15). This burst was detected by LHAASO (Large High Altitude Air Shower Observatory) in the 0.2–7 TeV range for $\sim$2000 s \citep{Cao23a}, with photons reaching $\sim$13 TeV \citep{Cao23b}. By comparing the expected cascade emission with Fermi-LAT (Large Area Telescope) upper limits, several recent works have derived IGMF lower bounds in the range $10^{-20}-10^{-16}$ G (\citealt{Huang23, Vovk24, Xia24, Dzhatdoev24, Mirabal23, davela26}.)}

The limits derived from GRBs are generally less constraining than those from AGNs (active galactic nuclei), mainly due to the transient nature of GRB VHE emission. Unlike AGNs, where the duration of the VHE activity is typically unknown and must be assumed, GRBs provide a directly measurable emission timescale. Their impulsive nature defines a clear 'time zero', enabling an accurate reconstruction of cascade photon arrival times relative to the primary emission.
Another advantage of using GRBs to constrain the IGMF is the role of plasma instabilities. 
Plasma instabilities have been proposed as a mechanism that could suppress cascade emission by rapidly dissipating the energy of electron-positron pairs (\citealt{Broderick12}), potentially preventing IGMF constraints. However, such instabilities require time to develop, and for transient VHE sources such as GRBs the growth timescale is expected to exceed the duration of the emission \citep{Batista21}.

All the results on GRBs discussed so far are based on searches for pair-echo emission in the GeV domain with Fermi-LAT. The upcoming Cherenkov Telescope Array Observatory (CTAO)  \citep{ctao} is expected to significantly enhance the sensitivity of pair-echo searches, extending the accessible energy range down to $\sim$20 GeV. This is particularly crucial for IGMF searches, as the reprocessed cascade signal is expected to peak in the tens-of-GeV regime.
Thanks to its improved performance compared to current imaging atmospheric Cherenkov telescopes (IACTs), CTAO can enable the search for pair-echo emission using the same instrument that detected the primary VHE photons. This avoids intercalibration issues between instruments and allows exploration of different IGMF configurations. The cascade time delay scales as $T_d \propto E^{-5/2} B^2$ for large correlation lengths with respect to the IC scale length \cite{Neronov09}, implying that higher-energy photons probe stronger magnetic fields.
\cite{Miceli23} investigated CTAO-North pair-echo detection prospects by simulating cascade emission from GRB 190114C and GRB 221009A, also exploring redshift dependence. They find that as little as 3 h of exposure after the afterglow phase could allow CTAO-North to detect or exclude certain IGMF configurations, with nearby GRBs being the most favourable targets.

In this work, we investigate IGMF constraints from GRB observations with CTAO by simulating time-dependent afterglow emission as a function of jet energy, redshift, and jet-break time. We compute the resulting pair-echo emission for different IGMF configurations and compare it with CTAO sensitivity to identify detectable regimes. We also assess the conditions under which the cascade emission can become comparable to or exceed the afterglow flux. The conclusions of this work are broadly applicable to other IACT facilities. However, current-generation instruments have significantly lower sensitivity than CTAO-North at $\sim$100 GeV, typically by a factor of $\sim$ 4. Among them, H.E.S.S. can reach an energy threshold as low as $\sim$30 GeV, but with a sensitivity degraded by a factor of $\sim$5–10 with respect to CTAO-North. MAGIC has a nominal threshold of $\sim$50 GeV, with a sensitivity nearly an order of magnitude lower than CTAO-North. For these reasons, we focus on CTAO-North as the reference instrument. Ground-based particle arrays such as LHAASO, with energy thresholds of $\gtrsim$300 GeV, are less suited for detecting GeV-scale pair-echo emission. However, their large duty cycle and wide field of view make them essential for the initial detection of VHE GRB emission and for triggering follow-up observations with IACTs.

The paper is structured as follows. In Section~\ref{sec:grb_igmf_models}, we present the GRB model adopted for this study and describe the parameter space explored. We also explain how we derive the pair-echo light curves for different IGMF configurations, taking into account the temporal evolution of the GRB afterglow emission in the VHE band. In Section~\ref{sec:results}, we present the results of the pair-echo calculations and discuss the conditions under which this emission might be detectable with CTAO. Finally, in Section~\ref{sec:conclusions}, we summarize the main results and outline future developments.

\section{GRB and IGMF models}
\label{sec:grb_igmf_models}
\subsection{GRB afterglow models}\label{sec:afterglow_model}
GRBs are produced by relativistic jets ejected from a central compact source generated following the core-collapse of a massive star or the merger of two neutron stars or a neutron star and a black hole. GRBs originated from the core-collapse of massive stars have a prompt emission lasting in general more than two seconds and are located at an average redshift $z\sim 3$ (long GRBs - \citealt{Ghirlanda2022}). As of today, five long GRBs have been clearly detected at VHE (from tens of GeV to $\sim$13\,TeV) in their afterglow phase, on time scales that range from tens of minutes up to a few days.
Afterglow emission from the radio band up to $\sim$\,GeV originates from synchrotron radiation emitted by electrons accelerated at the external shock running into the GRB surrounding medium (forward shock, {\citealt{meszaros93,Sari98,wijers99,ghisellini10,kumar10}}).
VHE radiation is instead interpreted as inverse Compton from the same electron population scattering off the synchrotron photons (synchrotron self-Compton process -- SSC, {\citealt{190114C_MAGIC_2019b,wang19,derishev21,joshi21,salafia22,barnard25}}).

Being relativistic sources ($\Gamma\sim10^2-10^3$), the observer only receives radiation coming from a region with a semiaperture angle $1/\Gamma$, which is initially smaller than the jet opening angle $\theta_{\rm jet}\sim$ a few degrees. When the shell decelerates down to $\Gamma\sim 1/\theta_{\rm jet}$ the observer receives radiation from the entire front of the jet and the flux starts to decay in time more rapidly. This produces a break in the afterglow lightcurves known as jet break time $t_b$, with typical values between a fraction of a day to a few days.

To explore the properties of the cascade and the prospects of detection as a function of the main properties of GRBs, we built simulated afterglow spectra as a function of time for different values of the initial isotropic-equivalent kinetic energy $E_{\rm k,iso}$ of the outflow, redshift $z$, and jet break time $t_b$. 
In particular, we considered thirteen values of $E_{\rm k,iso}$, equally spaced in logarithmic scale from $10^{49}$\,erg to $10^{55}$\,erg, sixteen values of $z$, equally spaced in logarithmic scale from 0.03 to 1, and four different jet-break times: $\log(t_{\rm b}\rm{[days]})=-1, -0.5, 0, 0.5$, in addition to the 'no jet break' scenario. 

For each pair of values ($E_{\rm k,iso}$, $z$), we simulated the synchrotron and SSC afterglow emission from electrons accelerated at the forward shock. The outflow was assumed to expand in a medium characterized by a constant number density $n$. A fraction $\xi_{\rm e}$ of the electrons swept up by the shock were accelerated and carry in total a fraction $\epsilon_{\rm e}$ of the shock energy. Their energy spectrum was assumed to be described by a power-law $dN/d\gamma\propto\gamma^{-p}$ from a minimum Lorentz factor $\gamma_{\rm min}$, determined by the amount of energy carried by the electron non-thermal population and their spectral index $p$. The generated magnetic field in the downstream was estimated assuming that it carries a fraction $\epsilon_{\rm B}$ of the shock energy. The evolution of the electron distribution and the resulting synchrotron and SSC emission were estimated accounting for adiabatic and radiative cooling, pair production and synchrotron self-absorption. {In this scenario, the peak energy of the SSC component ranges from 20 GeV to 20 TeV, with higher values preferred at earlier times.}
The details of the model and numerical implementation can be found in \cite{miceli_nava_review} and \cite{201216C}.

{To determine the values of the model parameters and predict the afterglow lightcurves and spectra, we adopted the method described in \cite{Ghirlanda2015}. In this approach, the afterglow emission from radio to TeV energies was simulated over a large grid of values for the unknown model parameters. The best fitting values were then identified by requiring the model to reproduce the observed distributions of X-ray and optical fluxes at 11\,h and radio fluxes at 2-5\,d (see \citealt{Ghirlanda2015} for further details). In addition, we required that the TeV simulated luminosity for the brightest bursts is comparable to the simultaneous X-ray luminosity, as inferred from the few GRBs detected so far at TeV energies. We find that a good description of multi-wavelength afterglow observations is obtained by adopting the following parameter values:
$n=0.5$\,cm$^{-3}$, $\xi_{\rm e}=0.2$, $p=2.3$,  $\epsilon_{\rm e}=0.04$, $\epsilon_{\rm B}=5\times10^{-5}$. We note that these values also correspond to typical values inferred from afterglow modelling. Once the model parameters have been determined,} the emission was simulated from 1 second to 7 days. Spectra were saved with a resolution in time equal to $\Delta \log(t) = 0.1$ up to 3 hours and every 3 hours at later times. The energy resolution of simulated spectra is $\Delta \log(E) = 0.1$. Finally, the jet break time and the evolution of the VHE lightcurve after $t_b$ were introduced manually, by steepening the light curve to include a variation in the temporal slope from $F_{\rm VHE}\propto t^a$ to $F_{\rm VHE}\propto t^{a-1}$ starting from $t_b$.

\subsection{Pair-echo modelling}

In order to model the pair-echo emission for each object in the ($E_{\rm k,iso}$, $z$, $t_{\rm b}$) grid (see Sect.~\ref{sec:afterglow_model}), we employed the Monte Carlo code CRPropa 3 (version 3.2; \citealt{Batista22}). This code allows us to trace the development of cascade emission in the intergalactic medium, given a certain VHE spectrum of a putative source located at a given redshift. In the simulations, the source was placed at the centre of a sphere with a radius equal to the source distance. Primary VHE photons were emitted within a cone of fixed aperture, and during their propagation pair production on the EBL and IC scattering processes were included. For the pair-production process, we adopted the EBL model of \cite{Franceschini08} as the target photon field. Each photon (either primary or cascade) that intersects the sphere was flagged as 'detected' and written to the output.
 
In our framework, $B$ was modelled as a zero-mean Gaussian turbulent field with a Kolmogorov spectrum and a correlation length of $\sim$ 5 Mpc, corresponding to the so-called large-correlation-length regime (with respect to the IC cooling length). As discussed in \cite{Neronov09}, lower bounds on the IGMF strength in this regime can be rescaled to the small-correlation-length regime following the relation $B \propto \lambda_{B}^{-1/2}$. The IGMF was defined in Fourier space, transformed into real space, and then projected onto a uniformly spaced cubic grid with $N=100^3$ cells of 50 Mpc each. This cubic cell was periodically repeated to fill the volume between the source and the observer. In this work, we tested the following physical rms field strengths: $B = 10^{-19}, 10^{-18}, 10^{-17}, 10^{-16}$ G. We injected $5 \times 10^5 $ primary VHE photons for each object defined in the ($E_{\rm k,iso}$, $z$) grid and for each tested $B$ value. All particles were propagated with a minimum integration step of $10^{-6}$ pc, corresponding to a temporal resolution of the order of 100 s. This value defines the time-delay precision adopted in the simulations. The adaptive integration tolerance was set to 10$^{-11}$. For each cascade photon, the simulations recorded its travel distance, which we used to compute the time delay $\tau$ relative to the source emission, as well as its energy $E$ and the energy of the primary VHE photon from which it originates $E_0$. In this way, we can reconstruct the pair-echo light curves within a given cascade-energy range.

Following the approach of \cite{Acciari23}, CRPropa3 provides the cascade signal $G(E_0,E,\tau, B, \lambda_{B})$ , namely the flux at energy $E$ produced by primary VHE photons of energy $E_0$, injected instantaneously by the source and arriving after a delay $\tau$. The cascade 'Green’s function' $G$ also depends on the IGMF configuration considered. In the present work, the primary source is a GRB afterglow characterized by a time-dependent VHE flux $F_s(E_0,t)$. The corresponding pair-echo flux $F_c(E,t)$ measured by the observer at time $t$ and energy $E$ can therefore be computed by convolving the cascade Green's function with the temporal evolution of the source emission:

\begin{equation}
F_c(E,t)=\int_{E}^{\infty}\int_0^{\infty} G(E_0,E,\tau)F_s(E_0,t-\tau)d\tau dE_0.
\end{equation}

The choice of the VHE spectrum injected into the intergalactic medium is a crucial aspect in properly estimating the pair-echo light curve. Our GRB simulations provide, for each time bin $t$, the corresponding VHE spectrum. To model the cascade, we adopted the following approach: we first derived the average spectrum 
$<F_E(E)>$ over the entire simulation time window $0-6.5\times10^5$ s. We then fixed the spectral shape to this average spectrum, while allowing the normalization to vary in time according to the VHE variability pattern $F_s(E_0,t)$. This is, of course, an approximation, but the procedure can be easily generalized in the future to consider different energy spectra in each time bin by including an additional convolution over the primary energy $E_0$.

To reduce the number of CRPropa simulations, we injected VHE power laws $dN/dE\propto E^{-2}$ in the energy range 100 GeV–100 TeV. Since, for each simulation, the primary energy $E_0$ of every cascade photon is recorded, the results can be straightforwardly rescaled to any desired custom VHE spectrum.

In Fig. \ref{fig:lc_plot} an example of pair-echo lightcurve for IGMF strength $B = 10^{-17}$ G and for the case $E_{\rm k,iso}=10^{53}$\,erg, $z=0.05$, and $\log(t_b/\rm{[days]})=-1$ is shown. We note that CRPropa3 is known to present limitations in the treatment of cascade propagation at high redshift and does not include the intrinsic cascade time delay (e.g. \citealt{Vovk23}, \citealt{Kalashev23}). For the present study, these effects are not expected to significantly affect the results. The pair-echo energies investigated here (20–125 GeV) are mainly produced by primary photons of a few–tens TeV, for which the optical-depth discrepancies remain of the order of 10$\%$ for a reference redshift of $z = 0.3$. In addition, the intrinsic cascade delays are typically $\lesssim$ 10$^3$-10$^4$ s in this energy range, whereas we focus on the detectable pair-echo signatures arising on timescales of $10^{4} - 10^{5}$ s or longer (see Fig. \ref{fig:lc_plot}). We therefore expect these limitations to have only a minor impact on our conclusions.

\section{Results}
\label{sec:results}
\subsection{Scan of GRB and IGMF parameters space}
\label{subsec:colorplots}

As explained in the previous section, we scan the parameter space defined by ($E_{\rm k,iso}$, $z$, $t_b$) to identify the GRBs whose physical properties make the pair-echo emission comparable to, or even dominant over, the afterglow emission. In addition, we test four different IGMF strengths to identify which are the most promising GRBs for each given configuration of the IGMF. 

\begin{figure}[!h]
\centering
\includegraphics[width=\columnwidth]{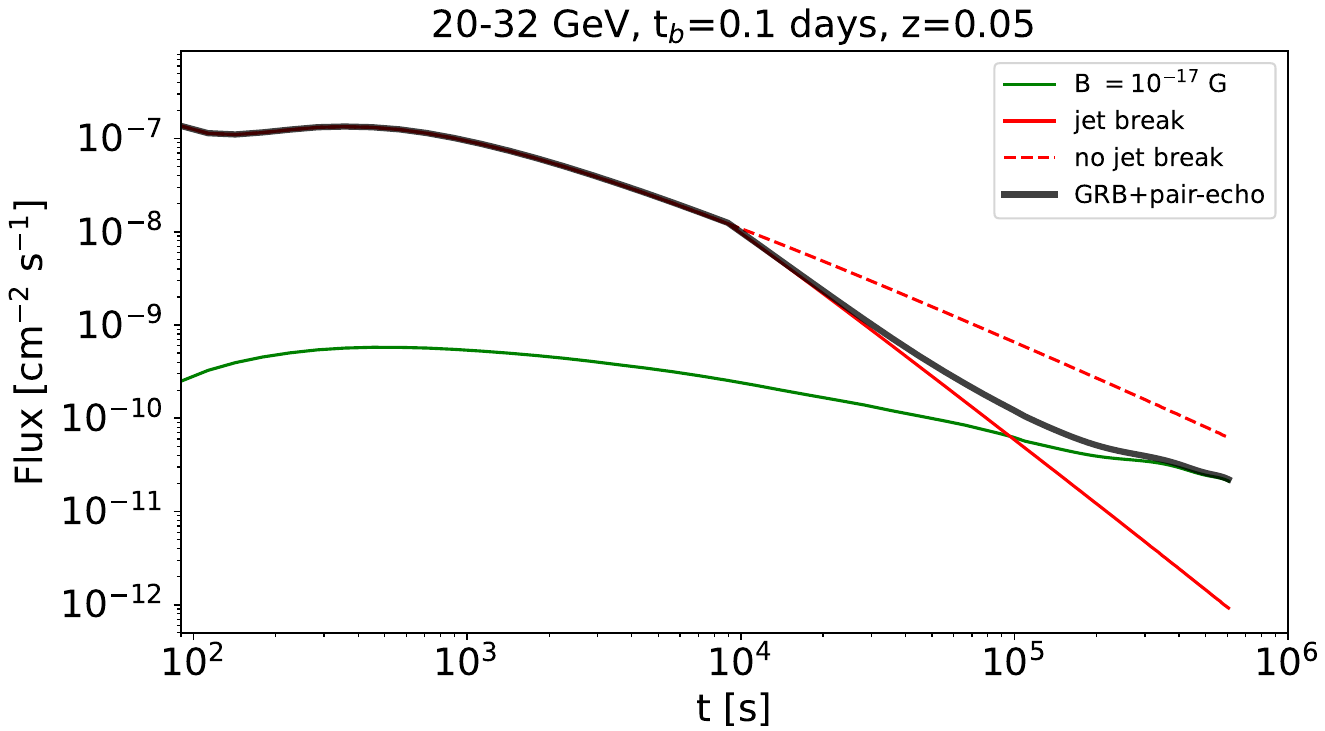} 
\caption{{Light curves in the 20 – 32 GeV energy range for a simulated GRB event with $z = 0.05$ and $E_{\rm k,iso} = 10^{53}$ erg. The GRB afterglow emission is shown in red lines in case of absence of a jet break (dashed line) or assuming a jet break steepening in the light curve at 0.1 days (solid line). The pair-echo emission calculated for the case including the jet break at 0.1 days for IGMF strength $B = 10^{-17}$ G is shown in green line. The sum of the two components (GRB afterglow and pair-echo) is shown in black line.}}
\label{fig:lc_plot}
\end{figure}

{An example of the expected light curve evolution for a set of values of $E_{\rm k,iso}$, $z$, $t_b$ and $B$ considering both the GRB afterglow (red lines) and the pair-echo emission (green line) is shown in Fig. \ref{fig:lc_plot}.} It is evident that one of the most crucial parameters is the GRB jet-break time (\citealp{Murase09}): indeed, the shorter this time, the higher the chances that the pair-echo light curve becomes comparable to—or even exceeds—the afterglow emission, making its detection possible. For simplicity we therefore present the results for the most favourable case, namely $\log(t_{\rm b}/\rm{[days]})=-1$. We notice that for GRB~221009A, a break in the TeV lightcurve at $\sim$670\,s is present and has been interpreted as a jet break time \citep{Cao23a}, thus corresponding to $\log(t_{\rm b}/\rm{[days]})\simeq-2$. Its interpretation is questioned by analyses of the X-ray and optical/near-infrared light curves \citep{Oconnor23}, which suggested the presence of a steepening at later times ($\log(t_{\rm b}/\rm{[days]})\simeq-0.1$), but in any case the TeV lightcurve shows a clear steepening. 
{Moreover, the cumulative emission resulting from the GRB afterglow and the pair-echo in the case of a jet-break at 0.1 days (black line in Fig. \ref{fig:lc_plot}) shows that when the pair-echo becomes dominant, a flattening of the cumulative light curve relative to the standard GRB afterglow evolution is expected. We estimate that the difference between the temporal indices of the cumulative light curve is $\Delta \alpha \approx 1$. Within the context of GRB physics, this feature cannot be explained by any effects due to the GRB afterglow emission and therefore constitutes a clear signature of pair-echo emission from GRBs.}

\begin{figure*}[!h]
\centering
\includegraphics[width=0.23\textwidth]{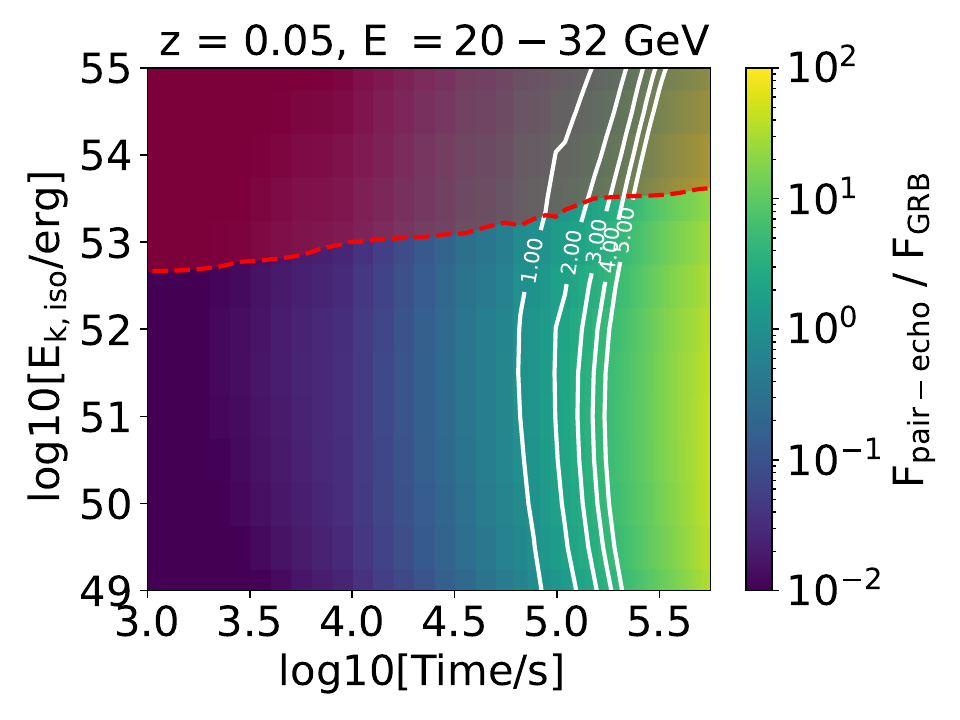} 
\includegraphics[width=0.23\textwidth]{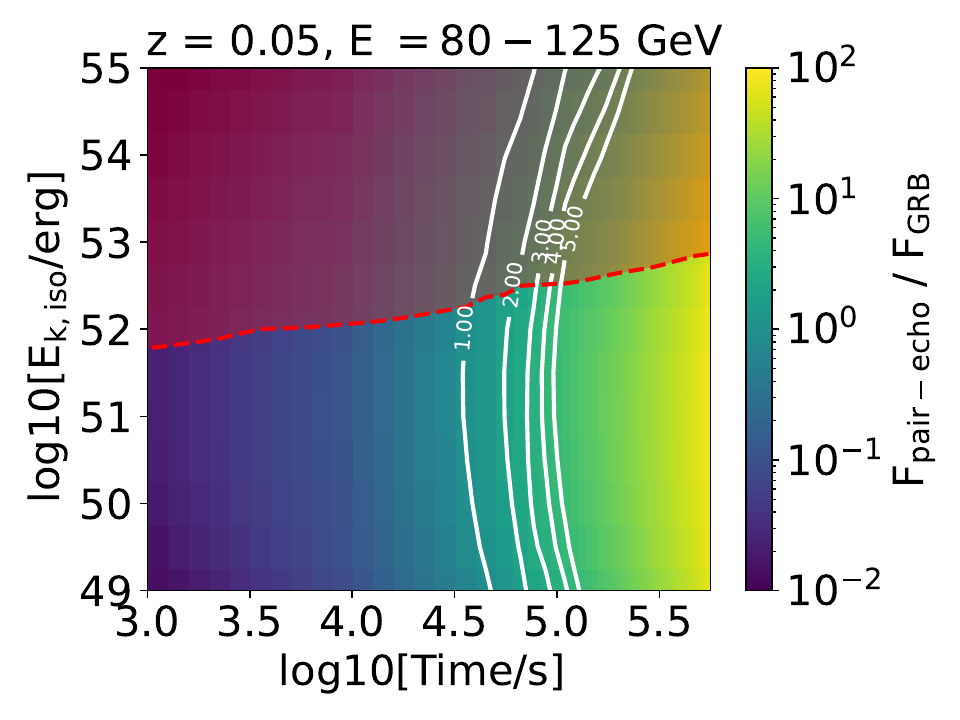}
\includegraphics[width=0.23\textwidth]{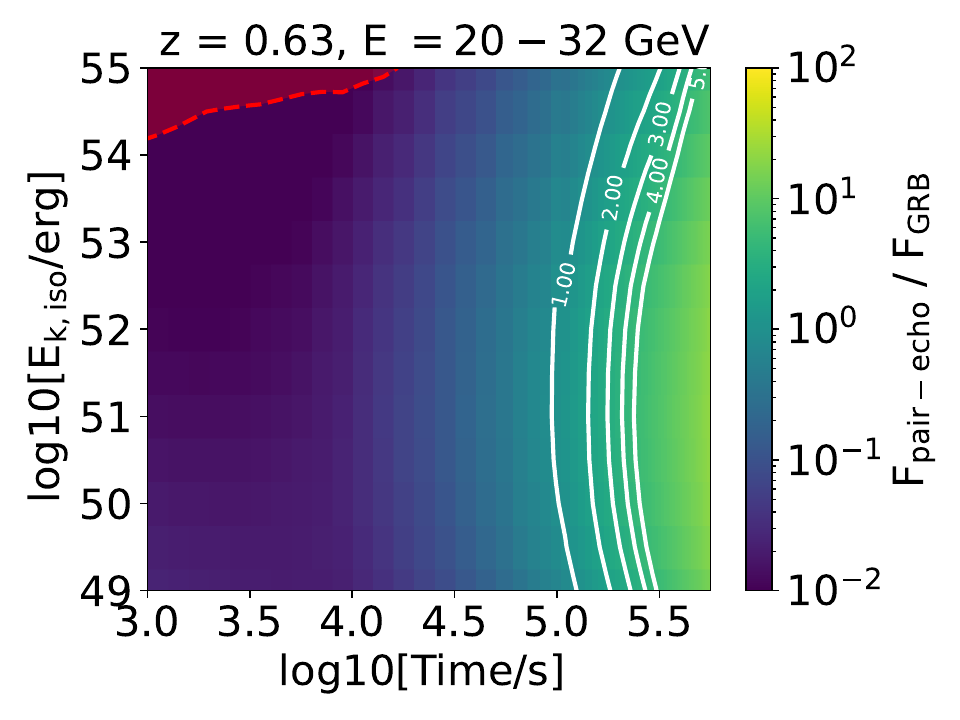}
\includegraphics[width=0.23\textwidth]{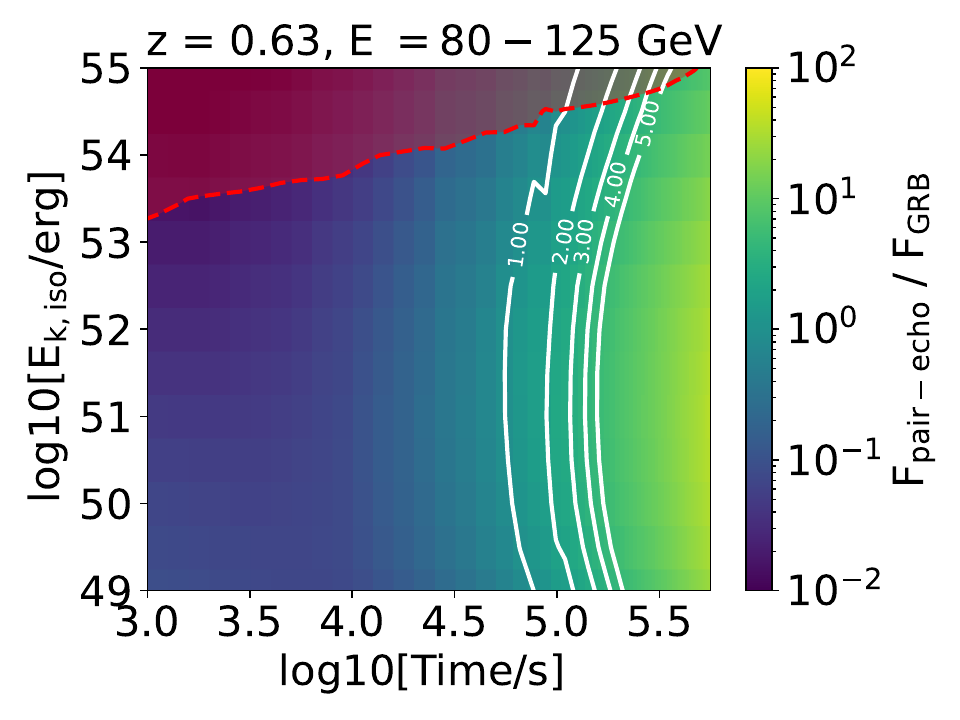}
\caption{Colour maps of the ratio between the pair-echo signal and the GRB afterglow flux ($F_{\text{pair-echo}}/F_{\text{GRB}}$) in the 20 - 32 GeV band (first and third plots from left)  and in the 80 - 125 GeV band (second and fourth plots from left) as a function of the observed time ($x$-axis) and $E_{\rm k,iso}$ ($y$-axis) in log scale. Plots are displayed for a set of fixed values for the IGMF strength $B = 10^{-17}$ G, redshift $z = 0.05, 0.63$ (from left to right) and assuming a jet break at $t_{\rm b} = 0.1$ days after GRB trigger time. Contour lines of the ratio ($F_{\text{pair-echo}}/F_{\text{GRB}}$) for values from 1 to 5 are displayed. The red shaded area represents the parameter space region where the pair-echo emission is detectable by CTAO-North following prescriptions described in Section \ref{subsec:colorplots}.}
\label{fig:colorplots_eiso_t}
\end{figure*}

To investigate the dependence on $E_{\rm k,iso}$, photon energy $E$, and redshift, we show in Fig. \ref{fig:colorplots_eiso_t} the colour maps of the ratio between the pair-echo and afterglow flux $F_{\rm pair-echo}/F_{\rm afterglow}$ as a function of the isotropic kinetic energy $E_{\rm k,iso}$ ($y$-axis) and time ($x$-axis)
for two energy bands, 20 - 32 GeV (first and third plots from left) and 80 - 125 GeV (second and fourth plots from left), and for two different redshifts, $z = 0.05$ (first and second plots from left) and $0.63$ (third and fourth plots from left). {We focus on these two energy bands because they probe complementary aspects of the pair-echo: the 20 –32 GeV band corresponds to a new low-energy GeV window accessible to CTAO, where most of the pair-echo emission is expected, while the 80–125 GeV band exploits the improved sensitivity of IACTs at higher energies, even though the cascade flux is expected to be lower.} In all these panels, the IGMF strength is $10^{-17}$ G and the jet break time is $t_b = 0.1$ days. The white contour lines mark the regions where the cascade flux equals or exceeds by factors of two, three, four, and five, the afterglow flux.

The first noticeable feature is a mild dependence of the time at which the cascade becomes dominant over the afterglow on $E_{\rm k,iso}$. From the shape of the contour lines, one can identify an 'optimal' range of $E_{\rm k,iso}$ where the pair-echo emission rises earlier. However, this dependence remains relatively weak. Another key feature concerns the dependence on the observed photon energy: for the energy bands considered in this work, the comparison between the 20–32 GeV and 80–125 GeV intervals shows that, for a fixed magnetic-field strength, the pair-echo component tends to overtake the afterglow at earlier times at higher observed photon energies. This behaviour is consistent with the expected dependence of the magnetic-field-induced cascade delay on energy, $T_d \propto E^{-5/2} B^2$ (in the large correlation length regime see, e.g. \citealt{Neronov09}). At energies approaching or exceeding the peak of the intrinsic SSC emission relevant for cascade production, the intrinsic source flux relevant for the cascade production decreases rapidly and the resulting cascade contribution is correspondingly reduced and may no longer be able to dominate over the afterglow. The relative evolution of the pair-echo and afterglow components is influenced not only by the magnetic-delay scaling but also by the shape of the intrinsic spectrum.

From Fig. \ref{fig:colorplots_eiso_t}, we also observe that increasing the redshift causes the contour lines to shift towards longer delays, indicating that the time at which the cascade becomes dominant occurs later for more distant GRBs. 

To assess whether the pair-echo emission can not only dominate over the afterglow emission but also be detectable by CTAO we consider the differential flux sensitivity of the CTAO Northern Array for five hours of exposure  for observations performed at a zenith angle of 20$^{\circ}$ at 20 - 32 GeV and 80 - 125 GeV ($F_{\rm CTAO}^{20\rm{GeV}}$ and $F_{\rm CTAO}^{80\rm{GeV}}$) taken from the official CTAO performance webpage \citep{CTAO2025}. In the currently approved alpha configuration, the Northern Array hosts the large-sized telescopes (LSTs, 23 m diameter), which lower the energy threshold of the system down to about 20 GeV. This is particularly relevant since the bulk of the pair-echo emission is expected to emerge in this energy range.

\begin{figure*}
\centering
\includegraphics[width=0.23\textwidth]
{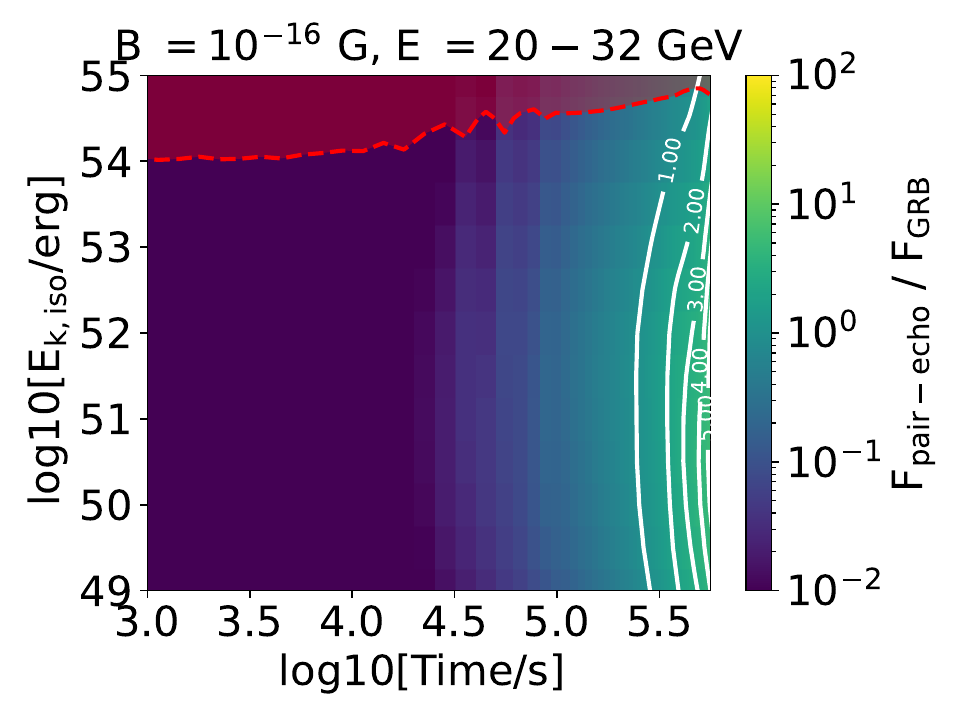} 
\includegraphics[width=0.23\textwidth]{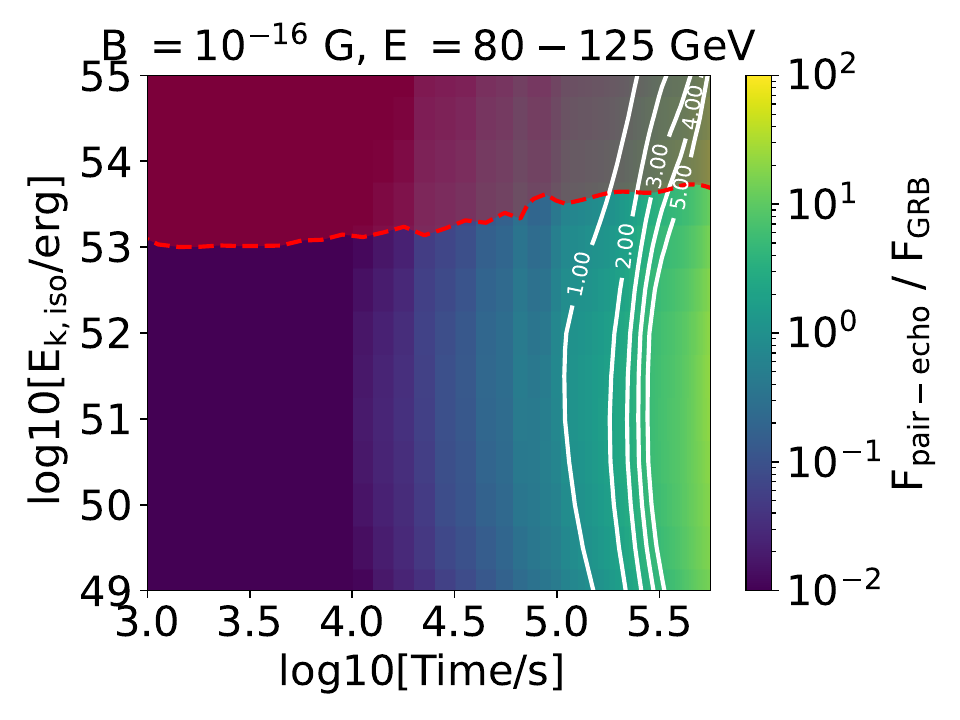} 
\includegraphics[width=0.23\textwidth]{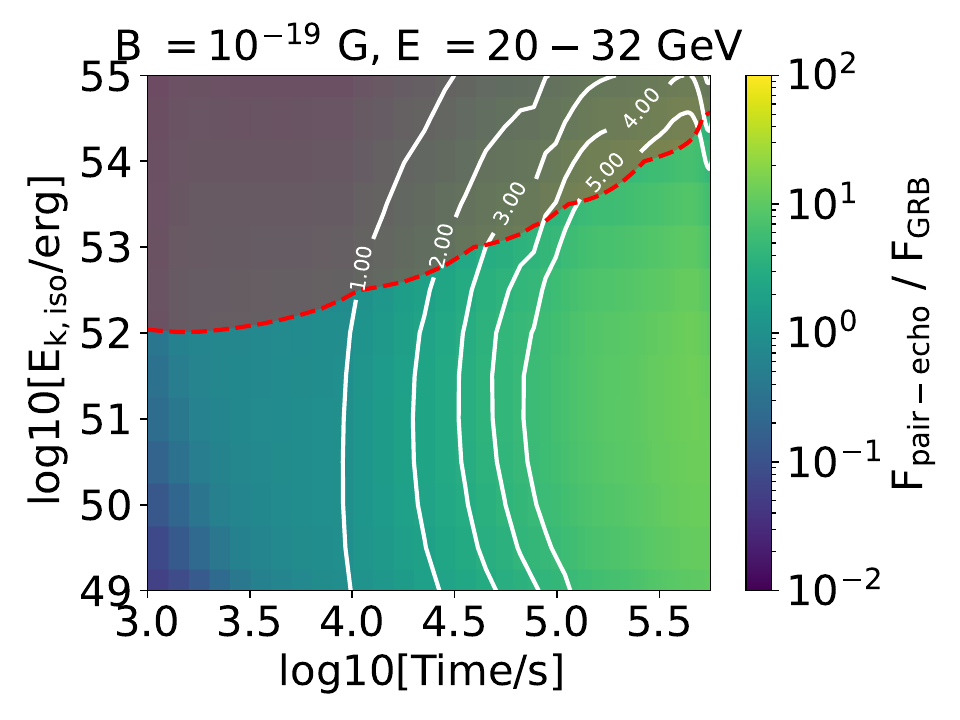}
\includegraphics[width=0.23\textwidth]{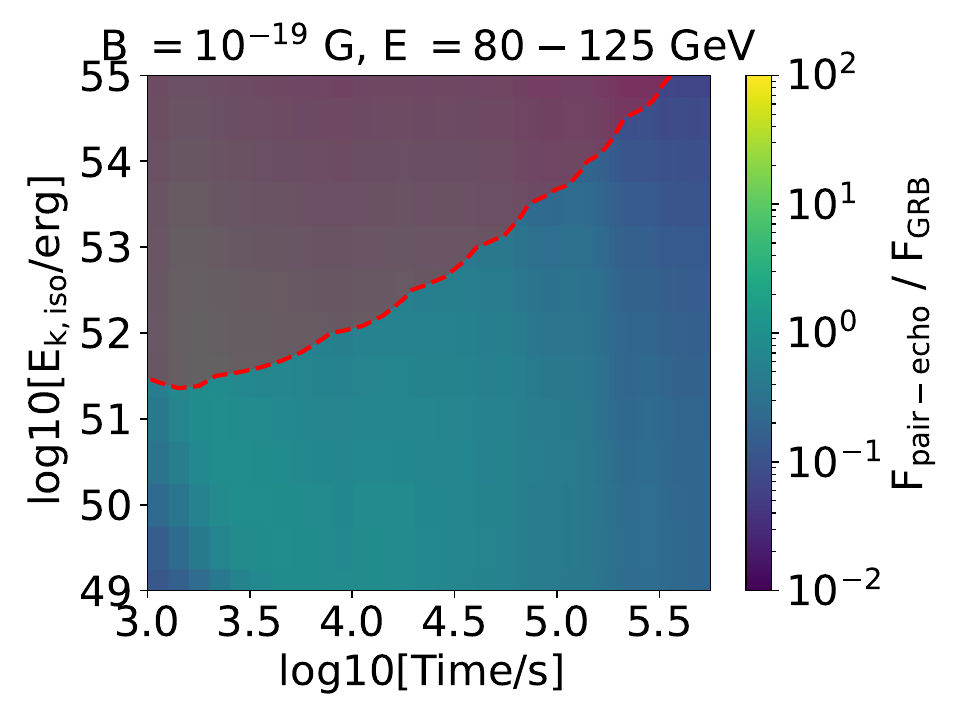}
\caption{Colour maps of the ratio between the pair-echo signal and the GRB afterglow flux ($F_{\text{pair-echo}}/F_{\text{GRB}}$)  in the 20 - 32 GeV band (first and third plots from left)  and in the 80 - 125 GeV band (second and fourth plots from left) as a function of the observed time ($x$-axis) and $E_{\rm k,iso}$ ($y$-axis) in log scale for the strongest and weakest IGMF strength values explored in this study, namely $B = 10^{-16}$ G and $10^{-19}$ G (first and second panels and third and fourth panels, respectively). Plots are displayed for a fixed value of redshift $z = 0.1$ and assuming a jet break at $t_{\rm b} = 0.1$ days after GRB trigger time. Contour lines of the ratio ($F_{\text{pair-echo}}/F_{\text{GRB}}$) for values from 1 to 5 are displayed. The red shaded area represents the parameter space region where the pair-echo emission is detectable by CTAO-North following prescriptions described in Section \ref{subsec:colorplots}.}
\label{fig:colorplots}
\end{figure*}

In Figs. \ref{fig:colorplots_eiso_t} and \ref{fig:colorplots}, the red-shaded regions of the colour maps indicate where $F_{\rm pair-echo} > F_{\rm CTAO}^{20\rm{GeV}}$ and $F_{\rm pair-echo} > F_{\rm CTAO}^{80\rm{GeV}}$,  for the left-hand and right-hand panels, respectively. As an example, when the white contour corresponding to $F_{\rm pair-echo} = 2 \times F_{\rm GRB}$ intersects the red region, the pair-echo emission would dominate the emission and be detectable by CTAO. 
The parameter space in which the cascade might be detectable is generally larger at the higher tested energy. Indeed, although the cascade flux is intrinsically higher in the 20 - 32 GeV band, the better sensitivity of the instrument in the 80 - 125 GeV band (approximately a factor $\sim 10$ better than at 20 - 32 GeV) plays a more significant role in determining the detectability of the emission. 

Fig. \ref{fig:colorplots} shows how the results depend on the IGMF strength, and allows to study which is the most suitable energy range to probe weak and strong IGMF strengths.  Similarly to Fig. \ref{fig:colorplots_eiso_t}, colour maps of $F_{\rm pair-echo}/F_{\rm afterglow}$ are shown as a function of $E_{\rm k,iso}$ and time. The first and third panels from left correspond to the 20–32 GeV band, while second and fourth panels correspond to the 80–125 GeV band. First and second panels refer to the strongest IGMF strength investigated ($B = 10^{-16}$ G) and third and fourth panels to the weakest ($B = 10^{-19}$ G). This time the redshift is the same in all panels and is $z = 0.1$ {and the jet break time is $t_b = 0.1$ days}. Owing to the dependence of the time delay on both the IGMF strength and photon energy (as discussed above), in the 80 - 125 GeV band the pair-echo emission associated with weak magnetic fields peaks at short delays, so that the cascade never becomes comparable to or dominant over the afterglow. In contrast, in the 20 - 32 GeV band, for the weakest tested magnetic field (B $= 10^{-19}$ G), there exist regions of the parameter space where the pair echo could be detectable by CTAO. Conversely, because of the better CTAO sensitivity at higher energies and the fact that, for stronger magnetic fields, the time delays are shorter in the 80 - 125 GeV band than in the 20 - 32 GeV band, the case B $= 10^{-16}$ G is more favourable in the 80 - 125 GeV band. In general, weak magnetic fields are better probed at lower energies, whereas strong magnetic fields can be more effectively investigated at higher energies.

\subsection{Detectable pair-echo events in the $z - E_{\rm k,iso}$  plane}
\label{subsec:stacked_analysis}
We focus on events fulfilling these criteria: (i) a pair-echo flux exceeding the GRB intrinsic emission by at least a factor 2; (ii) a pair-echo flux exceeding the CTAO reference thresholds $F^{20 \text{GeV}}_{\text{CTAO}}$ and $F^{80 \text{GeV}}_{\text{CTAO}}$. The condition (i) provides a conservative requirement to select cases in which the pair-echo component is sufficiently dominant over the afterglow to produce a clear flattening  in the total light curve. The choice of a factor of 2 is motivated by the fact that it exceeds the typical flux uncertainties achieved in current VHE GRB observations (see e.g. \citealt{HESS2021}), thereby reducing the risk of confusing the emergence of the pair-echo component with statistical fluctuations or uncertainties in the afterglow decay. The condition (ii) ensures that this flattening occurs while the flux is still detectable by CTAO.
With this sample, we can identify the regions of the GRB $z - E_{\rm k,iso}$ parameter space where the IGMF strength can be constrained. We considered results for $t_b = 0.1$ days. In addition, selecting two different energy bands (20 - 32 GeV and 80 - 125 GeV) we estimated the impact of the selected energy range for observations on the testable IGMF strength. 
For comparison with current IACTs, we also estimated events that exceed the MAGIC reference threshold in the 80 - 125 GeV energy range for five hours of exposure. The MAGIC sensitivity in this energy range was extracted from \citep{fioretti}.
Moreover, to show the impact of $t_b$ on the results, we report also the results for $t_b = 0.316$ days in the 20 - 32 GeV energy interval.

\begin{figure*}[!ht]
\centering
\includegraphics[width=0.48\textwidth]{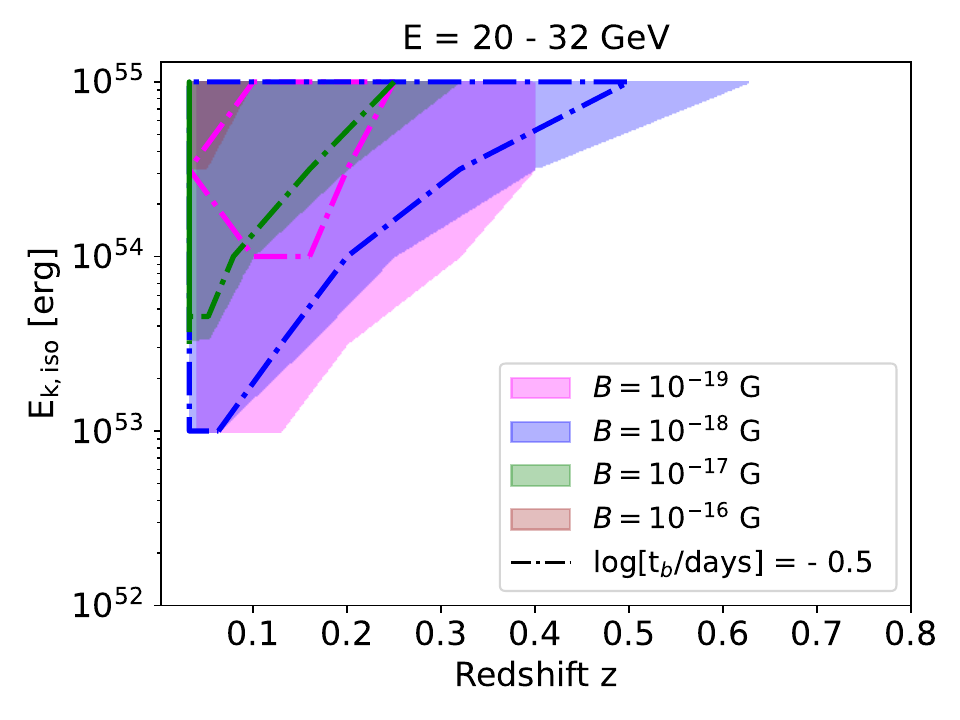} 
\includegraphics[width=0.48\textwidth]{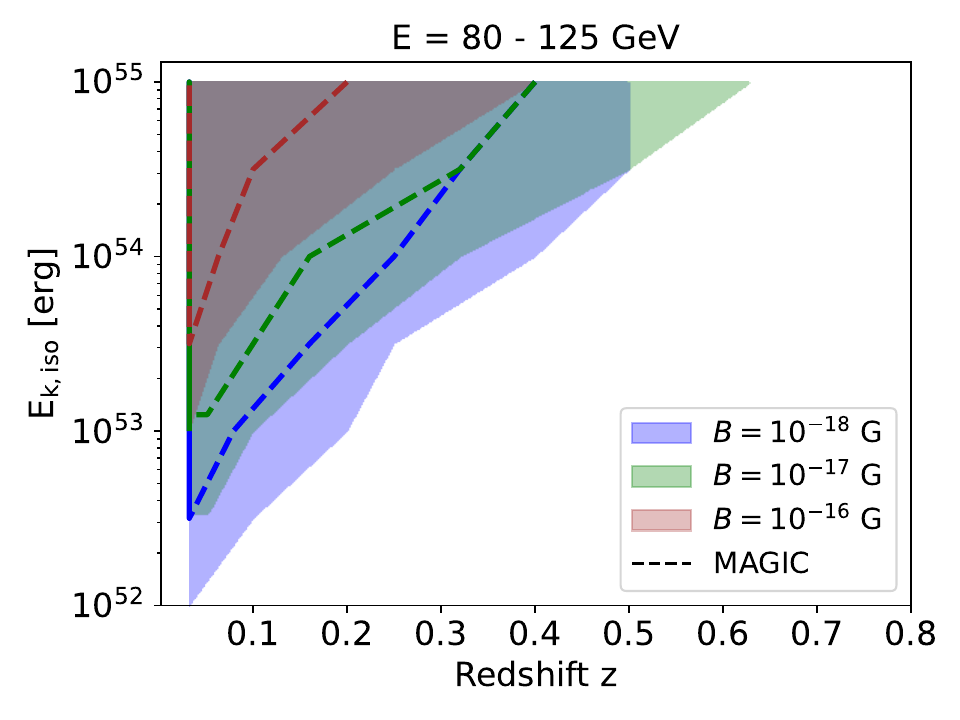}

\caption{Regions of the $z - E_{\rm k,iso}$ plane where the pair-echo signal is detectable with CTAO-North ({colour-filled regions}, see Section \ref{subsec:stacked_analysis}). Different colours correspond to different IGMF strengths. {Left-hand panel: photon energy range $ 20 - 32$ GeV; the regions corresponding to $t_b = 0.316$ days are also shown as dash-dotted contours. Right-hand panel: $80 - 125$ GeV. The regions detectable by MAGIC are overplotted as dashed contours.}}
\label{fig:stack_overplot_b}
\end{figure*}

In Fig. \ref{fig:stack_overplot_b} we display the regions of the $z - E_{\rm k,iso} $ parameter space that fulfil the two conditions mentioned above for each of the tested values of the IGMF strengths and for observations performed in the 20 - 32 GeV band (left plot) and 80 - 125 GeV band (right plot). Considering the total filled envelope covered by those regions, we conclude that only GRBs with $E_{\rm k,iso} \gtrsim 10^{52} - 10^{53}$ erg are suitable for IGMF studies. In addition, the limiting horizon for IGMF studies with GRBs is reduced to only nearby events of $z < 0.2$ for GRBs with $E_{\rm k,iso} < 10^{53}$ erg but it increases exponentially with $E_{\rm k,iso}$ reaching a limit of $z \sim 0.5 - 0.6$ for very bright events of $E_{\rm k,iso} = 10^{55}$ erg. No evident differences can be seen between the two selected energy ranges of 20 - 32 GeV and 80 - 125 GeV even though the latter one seems to be sensitive for events below $10^{53}$ erg and has a slightly larger limiting horizon, likely due to the better sensitivity of the CTAO telescopes at this energy.

For each given value of $B$, a clear difference is present between $20 - 32$\,GeV and $80 - 125$\,GeV, indicating that different observed energies provide different discrimination power on the IGMF strength. In particular, in the 20 - 32 GeV band, weaker IGMF strengths of $10^{-17} - 10^{-19}$ G can be tested in several combinations of $E_{\rm k,iso}$ and $z$, while for $10^{-16}$ G only the brightest ({$E_{\rm k,iso} > 10^{54}$} erg) and nearby ($z < 0.15$) GRBs can provide valuable IGMF studies. On the other hand, in the 80 - 125 GeV band, stronger IGMF strength in the $10^{-16} - 10^{-18}$ G can be probed covering almost the entire allowed parameter space for $E_{k,iso} > 10^{52}$ erg and $z < 0.5 - 0.6$. In addition, it can be seen that the IGMF configuration with $B = 10^{-19}$ G cannot be tested at these energies. This can be explained considering the combination of the short delays induced for secondary events in the 80 - 125 GeV and for such a weak IGMF strength.

{In the right plot of Fig. \ref{fig:stack_overplot_b} (80 - 125 GeV), we can compare the regions of the $z - E_{\rm k,iso} $ parameter space detectable by MAGIC (dashed contours) and CTAO (colour-filled regions). It is evident that the better sensitivity of CTAO increases the regions both in terms of limiting horizon $z$ and in terms of energetics $E_{\rm k,iso}$. Nevertheless, a portion of the brightest and nearby events with $E_{\rm k,iso} > 10^{53}$ erg and $z < 0.4$ can also be detectable by MAGIC. CTAO detectable regions for different IGMF strengths $B$ are larger than MAGIC ones by a factor $\sim 2$ in terms of redshift and by a factor $\sim 3$ in terms of energetics.
In the left plot of Fig. \ref{fig:stack_overplot_b} (20 - 32 GeV), we also include the regions of events detectable by CTAO assuming $t_b = 0.316$ days (dash-dotted contours). When comparing these results with those obtained for $t_b = 0.1$ days (colour-filled regions) we estimate that the number of detectable events in our grid decreases by $\sim 35\% $.
}
Another parameter that can provide valuable information, both from theoretical and observational perspectives, is the time at which the pair-echo signal becomes dominant and detectable. 
The estimates of the minimum time when the pair-echo emission overshoots the afterglow emission and becomes detectable by CTAO are reported in Tables \ref{tab:eiso_time_20} and \ref{tab:eiso_time_100} for different combinations of $E_{\rm k,iso}$, $z$, $B$ and for the two selected energy ranges. Also in this case, we considered only the case of a jet break appearing at $t_{\rm b}= 0.1$ days. As shown in Fig. \ref{fig:lc_plot}, the presence of a jet break plays a crucial role in anticipating the time when the cascade emission dominates over the afterglow radiation.

By inspecting the tables, it is evident that the pair-echo emission starts to dominate only at late times, from $\gtrsim 10-12$ h after the GRB explosion. The weaker IGMF strengths of $10^{-18}$ G and $10^{-19}$ G can be probed starting from $\sim 8-15 $ h after the burst trigger. A preference for the energy range 20 -  32 GeV is present for such IGMF configurations. On the other hand, for stronger $B$ ($10^{-16}-10^{-17}$ G) the pair-echo emission arises later, from $\sim 18$ h up to $ > 4$ days after the burst trigger. These IGMF strengths are likely detectable in the 80 - 125 GeV band rather than at lower energies. These results give a clear indication that late-time (from hours to days) observations of GRBs can provide valuable results to search for IGMF strengths. In addition, depending on the delay when the follow-up is performed, different IGMF strengths from $10^{-16}$ G up to $10^{-19}$ G can be tested. As a result, the limited available observational windows of Cherenkov telescopes should be tuned on the IGMF strengths to be tested. A clear evaluation of this aspect requires the usage of instrument response function of the telescopes. This will be part of a dedicated study in the next future. In conclusion, Cherenkov telescopes have strong capabilities in detecting the pair-echo emission from GRBs, despite observational limitations due to their low duty cycle. We note that the pair-echo light curve also depends on the IGMF correlation length. In this work we fixed $\lambda_{B}$ to the large-correlation-length regime, so that our results should be interpreted as a detectability study for this reference configuration. Reducing the correlation length shifts the cascade emission towards shorter delays (see e.g. \citealt{Neronov09}). Depending on the magnetic-field strength, the correlation length, and the temporal evolution of the GRB emission, this may either enhance the pair-echo flux at the observation times considered (\citealt{Vovk24}) or shift a significant fraction of the cascade emission to earlier times. A systematic exploration of the dependence on the correlation length therefore requires dedicated simulations over the full (B,$\lambda_{B}$) parameter space.

\section{Conclusions}
\label{sec:conclusions}
In this work, we present a comprehensive study on IGMF signatures from GRBs in the tens to hundreds of GeV energy range. In particular, we focus on the time-delayed pair-echo signature. We explore the capability of CTAO to detect these signals exploiting a sample of simulated GRB afterglows. We investigated how properties of GRBs, such as their energetics and distance, affect the possibility to detect the pair-echo emission during the afterglow phase, by exploring different IGMF configurations. In particular, we scanned the GRB parameter space defined by the outflow (isotropic equivalent) kinetic energy $E_{\rm k,iso}$ during the afterglow phase, the redshift $z$, and the jet break time $t_b$. We computed, for each set of parameters, the VHE lightcurve under the assumption of SSC emission, and the expected pair-echo light curve for various IGMF strengths, taking into account the temporal evolution of the GRB afterglow flux in the VHE band. These calculations were performed in two narrow energy bands, namely $20 - 32$ GeV and $80 - 125$ GeV, selected as a compromise between the range where the bulk of the cascade emission is expected and the energy domain in which the next-generation Cherenkov facility, CTAO, will operate.

From this work, we conclude that GRB emission in the VHE domain can produce non-negligible IGMF signatures, at least in a subclass of events. In addition, CTAO has the potential to detect and possibly disentangle, for the first time, both the afterglow GRB emission and the associated pair-echo signal. The combination of its broad energy coverage (20 GeV–300 TeV) and significantly improved sensitivity compared to current IACT facilities make it possible to probe different IGMF configurations by exploiting the dependence of the pair-echo flux on both energy and magnetic field strengths.

This investigation explores the detectability of the pair-echo signal as a function of key GRB parameters and demonstrates that a detection is possible for a subclass of events. The estimate of the detection performances (e.g. detection rate) of CTAO requires the implementation of a realistic GRB cosmic population and additional features such as the jet orientation with respect to the line of sight. This aspect is left for a follow-up study.

The main results of this study can be summarized as follows:

\begin{itemize}

\item by requiring that the cascade flux exceeds at least twice the afterglow emission and that it lies above the CTAO sensitivity threshold, we find that GRBs with isotropic kinetic energies $E_{\rm k,iso} \gtrsim 10^{52} - 10^{53}$ erg are suitable for IGMF studies.

\item The redshift range that can be probed depends on the kinetic energy $E_{\rm k,iso}$.
Lower values of E$_{\rm k,iso}$ limit the observable horizon, while the highest tested energies allow us to reach redshifts up to $z\sim$ 0.5–0.6.

\item The region of the ($E_{\rm k,iso}$, $z$) parameter space that can be probed also depends on the energy $E$ at which the pair-echo emission is observed. In general, the accessible region is broader at $80 - 125$ GeV than at $20 - 32$ GeV. This is mainly due to the better CTAO sensitivity in the former energy range compared to the latter one, even though a larger fraction of the cascade emission is expected at lower energies.

\item Due to the dependence of the time delay on both the cascade energy and the magnetic field strength, IGMF strengths in the range $10^{-19}-10^{-17}$ G can be probed over a wide region of the ($E_{\rm k,iso}$, $z$) parameter space in the $20 - 32$ GeV energy range. Stronger magnetic fields, instead, can be investigated only for nearby and bright GRBs. Conversely, in the $80 - 125$ GeV energy range, IGMF strengths around $10^{-16}$ G can be explored across a broad portion of the parameter space. Although fields of 
$10^{-18}-10^{-17}$ remain accessible at these energies, weaker magnetic fields cannot be effectively probed.

\item The time at which the pair-echo emission overshoots the afterglow strongly depends on the jet break time: the earlier the jet break occurs, the sooner the cascade can emerge. This implies that the pair-echo signal is detectable in case of GRBs with narrow jets (as in the case of GRB 221009A) or with steepening of the GRB afterglow light curve. In the most favourable scenario assuming $t_b = 0.1$ days, it corresponds to a range $\theta_{\rm jet}\in[1.5,3.0]^{\circ}$. As a result, weak magnetic fields can be probed within the first 10–12 hours after the trigger, while for stronger magnetic fields the pair-echo dominance occurs on longer timescales, from several tens of hours up to a few days. The results obtained assuming $\log(t_b[\rm days]) = -0.5,0,0.5 $ days follow the same trend, but the rising of the pair-echo component occurs for longer delays and in a reduced number of bright and nearby events.
\end{itemize}

\begin{acknowledgements}
LN and GG acknowledge funding by the European Union-Next Generation EU, PRIN 2022 RFF M4C21.1 (202298J7KT - PEACE). GG and LN acknowledge the project GRB PrOmpt Emission Modular Simulator (POEMS) financed by INAF Grant 1.05.23.06.04. DM, GG and LN acknowledge support from the Italian Space Agency,  through contract ASI n. 2025-5-U.0: 'Gamma-ray bursts: a probe of multi-messenger and extreme Astrophysics'. D.M. and P.D.V. acknowledge 'funding by the European Union – NextGenerationEU' RFF M4C2 project IR0000012 CTA+.
\end{acknowledgements}

\bibliographystyle{aa} 
\bibliography{biblio.bib}

\clearpage
\onecolumn
\begin{appendix}
\section{Pair-echo detection times}
\begin{table}[!ht]
\caption{Times at which the pair-echo flux becomes larger than the GRB afterglow flux and is detectable by CTAO.}\label{tab:eiso_time_20}
\centering
\begin{tabular}{cc|cccc}
\hline
\multicolumn{6}{c}{$E = 20 - 32$ GeV} \\
\hline
  $E_{\rm k,iso}$ & $z$ & $B=10^{-16}$ G & $B=10^{-17}$ G & $B=10^{-18}$ G & $B=10^{-19}$ G \\
 \tiny{erg} & & &&& \\
\hline
\hline
 $10^{53}$ & $0.05$ & - & - & $12.5$ h  & $18.4$ h \\
 $3.16 \times 10^{53}$ & $0.05$ & - & $36.4$ h & $12.5$ h  & $18.4$ h \\
 $3.16 \times 10^{53}$ & $0.13$ & - & - & $15.4$ h  & $7.9$ h \\
 $10^{54}$ & $0.08$ & - & $42.4$ h & $12.5$ h  & $9.9$ h \\
  $10^{54}$ & $0.13$ & - & - & $15.4$ h  & $7.9$ h \\
 $10^{54}$ & $0.2$ & - & - & $15.4$ h  & $9.9$ h \\
 $3.16 \times 10^{54}$ & $0.05$ & $159$ h & $39.4$ h & $12.5$ h  & $12.5$ h \\
 $3.16 \times 10^{54}$ & $0.1$ & - & $45.4$ h & $15.4$ h  & $7.9$ h \\
 $3.16 \times 10^{54}$ & $0.2$ & - & $54.4$ h & $15.4$ h  & $9.9$ h \\
  $3.16 \times 10^{54}$ & $0.4$ & - & - & $21.4$ h  & $15.4$ h \\
 $10^{55}$ & $0.1$ & $150$ h & $45.4$ h & $15.4$ h  & $7.9$ h \\
 $ 10^{55}$ & $0.25$ & - & $54.4$ h & $18.4$ h  & $9.9$ h \\
 $10^{55}$ & $0.4$ & - & - & $21.4$ h  & $15.4$ h \\
\hline
\end{tabular}
\tablefoot{Estimates were performed for E $= 20 - 32$ GeV and for different combinations of $E_{\rm k,iso}$, $z$ and $B$, and an early jet break $t_{\rm b}= 0.1$ days.
}
\end{table}

\begin{table}[h]
\caption{Times at which the pair-echo flux becomes larger than the GRB afterglow flux and is detectable by CTAO.}\label{tab:eiso_time_100}
\centering
\begin{tabular}{cc|cccc}
\hline
\multicolumn{6}{c}{$E = 80 - 125$ GeV} \\
\hline
 $E_{\rm k,iso}$ & $z$ & $B=10^{-16}$ G & $B=10^{-17}$ G & $B=10^{-18}$ G & $B=10^{-19}$ G \\
 \tiny{erg} & & &&& \\
\hline
\hline
  $10^{52}$ & $0.032$ &  - & - & $9.9$ h & - \\
$3.16 \times 10^{52}$ & $0.05$ &  -  & $18.4$ h & $9.9$ h & - \\
$3.16 \times 10^{52}$ & $0.1$ &  -  & - & $9.9$ h & - \\
$10^{53}$ & $0.05$ &  - & $18.4$ h & $9.9$ h & - \\
$10^{53}$ & $0.1$ &  - & $21.4$ h & $9.9$ h & - \\
$10^{53}$ & $0.2$ &  - & - & $12.5$ h & - \\
$3.16 \times 10^{53}$ & $0.05$ &  $66.4$ h  & $18.4$ h & $9.9$ h & - \\
$3.16 \times 10^{53}$ & $0.1$ &  - & $21.4$ h & $9.9$ h & - \\
$3.16 \times 10^{53}$ & $0.2$ &  - & $27.4$ h & $12.5$ h & - \\
$10^{54}$ & $0.08$ &  $81.4$ h  & $21.4$ h & $9.9$ h & - \\
$10^{54}$ & $0.16$ &  -  & $24.4$ h & $12.5$ h & - \\
$10^{54}$ & $0.32$ &  -  & $30.4$ h & $21.4$ h & - \\
$3.16 \times 10^{54}$ & $0.1$ &  $75.4$ h  & $21.4$ h & $9.9$ h & - \\
$3.16 \times 10^{54}$ & $0.25$ &  $90.4$ h  & $30.4$ h & $15.4$ h & - \\
$3.16 \times 10^{54}$ & $0.5$ &  -  & $42.4$ h & $42.4$ h & - \\
$10^{55}$ & $0.2$ &  $87.4$ h  & $27.4$ h & $12.5$ h & - \\
$10^{55}$ & $0.4$ &  $111$ h  & $39.4$ h & $30.4$ h & - \\
$10^{55}$ & $0.5$ &  -  & $42.4$ h & $30.4$ h & - \\
$10^{55}$ & $0.63$ &  -  & $57.4$ h & $42.4$ h & - \\
 \hline
\end{tabular}
\tablefoot{Estimates were performed for E $= 80 - 125$ GeV and for different combinations of $E_{\rm k,iso}$, $z$ and $B$, and an early jet break $t_{\rm b}= 0.1$ days.
}
\end{table}
\end{appendix}

\end{document}